\documentclass[epj]{svjour}
\usepackage{graphicx}
\usepackage{amsmath}
\def\slaninafigdir{.}
\begin{document}
\title{%
Self-organized branching process for a one-dimensional ricepile model%
}
\author{%
Franti\v{s}ek Slanina%
}
\institute{%
        Institute of Physics,
	Academy of Sciences of the Czech Republic,\\
	Na~Slovance~2, CZ-18221~Praha,
	Czech Republic\\
        e-mail: slanina@fzu.cz
}%
\date{%
(This version was processed on: \today )
}
%
\abstract{%
A self-organized branching process is introduced to describe
one-dimensional ricepile model with stochastic topplings.
Although the branching processes are generally supposed to describe
well high-dimensional systems, our modification grasps some of
the peculiarities present in one dimension.
We find analytically the crossover behavior 
from the trivial one-dimensional BTW
behaviour to self-organized criticality characterised by power-law
distribution of avalanches. The finite-size effects, which are crucial
in the crossover, are calculated.
} 
\maketitle
\section{Introduction}
Since the pioneering work of Bak, Tang and Wiesenfeld (BTW)
\cite{ba_ta_wi_87,ba_ta_wi_88}, the sandpile model became one of
prototype abstract 
models exhibiting self-organized criticality (SOC). The original BTW model 
as well as its variants
(see e.g. \cite{ka_na_wu_zho_89,gra_ma_90,manna_91,manna_91a,che_sta_ve_za_98})
consists of a cellular automaton slowly driven by stochastic
perturbation. The state of each site is described by number of grains
on top of it. (Actually, this number is rather the slope than the
height, if we like to interpret the model as a real
sandpile. However, in 1D models, investigated here, the description
through slope and height variables are strictly equivalent.)
If the number of grains exceeds a threshold, the site
becomes active, a toppling occurs and grains are transferred to
neighbouring sites, which then may become active and the process
continues. The driving consists in adding grains on randomly
chosen sites.
The critical state is reached asymptotically in the
limit of infinitely slow driving \cite{so_jo_do_95}.
Fully deterministic versions were also studied, showing periodic
\cite{wi_the_na_90,ma_ma_92}
or self-similar but non-random behaviour \cite{he_cha_de_ro_wa_99}.

Even though experiments on real sandpiles did not confirm SOC behaviour,
due to inertia effects 
\cite{ja_li_na_89,rajchenbach_90,he_so_ke_ha_ho_gri_90,nagel_92,pra_ola_92,ba_me_96,ba_wo_96}, 
in  the experiments using rice 
\cite{fre_chri_ma_fe_jo_mea_96,ma_fe_chri_fre_jo_99}
instead of sand it was found that large
aspect ratio of the rice grains can lead to SOC behaviour
\cite{fre_chri_ma_fe_jo_mea_96},  contrary to the case of sand, which
has grains much closer to spherical. 

Another difference between a typical sandpile and ricepile experiments
is that the ricepiles used in the experiments are 
quasi one-dim\-ensional \cite{fre_chri_ma_fe_jo_mea_96,ma_fe_chri_fre_jo_99}. 
While the original BTW model in one
dimension is trivial, there are several variants of 1D BTW model
which exhibit non-trivial behaviour
\cite{ka_na_wu_zho_89,he_cha_de_ro_wa_99,me_ba_94,sorensen_96,lu_us_98,di_al_mu_pe_ve_za_01,pri_iva_pov_hu_01%
}.
Also the sandpiles on quasi one-dimensional stripes were investigated
\cite{ma_ta_zha_99}.
 Several one-dimensional models devised especially for modelling the
ricepiles were studied
\cite{ama_la_96,ama_la_96a,ama_la_97,chri_co_fre_fe_jo_96,pa_bo_96,sdzhang_97,be_be_mhi_zha_99,be_be_ke_lo_mhi_99,ma_je_la_sne_97,markosova_00,sdzhang_00}.
The models taking into account a possible long-range rol\-ling of grains
are able to describe the transition from SOC behaviour typical for
ricepiles to the inertia-dominated behaviour of sand heaps
\cite{gle_can_tam_zhe_01,gleiser_01}.

Besides numerous exact results and renormalisation-group calculations
(to cite only a few items of a vast bibliography, see
\cite{dhar_89,dha_ma_90,markosova_95,klo_mas_tan_01,pie_ve_za_94,ivashkevich_96,zhang_89a}), 
the mean-field approximation
\cite{ta_ba_88a,ka_ko_96,ve_za_97a}
was very useful in clarifying the nature of the SOC state, even though
it cannot give correct values of the exponents below the upper
critical dimension.

It was realised soon that the mean-field approximation for sandpiles
is related to critical branching processes \cite{alstrom_88,garcia_94}. 
This idea lead to the introduction of 
self-organized branching processes
\cite{za_la_sta_95,la_za_sta_96,ca_te_ste_96,ve_ma_ba_97,chu_ada_99,chu_ada_99a},
which describe the approach to the critical state.
Similar approach consist in mapping the sandpile to the percolation on
Bethe lattice \cite{so_va_an_99}.

The approximation is based on the observation that in high dimension
activity returns to the same site with very small probability. So, we
can suppose that in each step the toppling occurs at a site, which
never toppled before during the same avalanche. 
Each toppling is mapped to one branching.
Statistical properties
of avalanches are determined by the probability $p$ of branching. This
probability is itself determined self-consistently. If the avalanche
is sub-critical, it does not fall off the system and average number of
grains, and thus $p$, increases. If, on the other hand, the avalanche
is super-critical, it surely falls off the system, which leads to
decrease of the average number of grains and decrease of $p$. It was
shown \cite{za_la_sta_95}, that this process sets the $p$ exactly to
the critical value, 
where the avalanche sizes $s$ have power-law distribution $P(s)\sim s^{-\tau}$ with mean-field
exponent $\tau=\frac{3}{2}$.

The purpose of this work is to modify the self-organized branching
processes in order to describe one-dimensional ricepile
models. Our model will be designed to comprise the one-dimensional BTW
model as a special case. Clearly we cannot obtain correct values of
the exponents. Our main question will be, whether there is a sharp
transition from trivial 1D BTW behaviour to SOC behaviour
or what is the nature of the crossover from the former to the latter.

The paper is organised as follows.  
In the next section we define our version of the branching process,
suitable for treating the one-dimensional ricepile. We find the
condition for the criticality and investigate the crossover from the
trivial one-dimensional BTW behaviour to the critical branching process.
The self-organization toward the critical state is investigated in
the section 3. We first define the self-organized branching process,
then find the fixed point of the dynamics and show that it exactly
corresponds to critical branching process. We finally investigate the
influence of finite size effects and find the finite-size scaling
form. The section 4 concludes and summarises the work.

\section{Branching process for one-dimensional model}

\subsection{Ricepile model}
The ricepile models were already thoroughly investigated by
numerical simulations. 
In fact, there are two variants of the one-dimensional ricepile
model. The so-called ``Oslo model'' 
\cite{chri_co_fre_fe_jo_96,pa_bo_96,sdzhang_97,be_be_mhi_zha_99}
supposes that the critical slope depends on space and time, and
assumes new random value after each toppling event. Another approach
\cite{ama_la_96,ama_la_96a,ama_la_97} assumes that the toppling occurs
with certain probability, which depends on actual slope. It is the
second approach, which we will follow in this article. 
It may be also noted that a two-dimensional model which also implements
stochastic topplings was studied before \cite{ou_lu_di_93}. 

We recall shortly the definition of the model. 
We consider a chain of $L$ sites. The state of site $i$, $i=1,2,...,L$
is described by a slope $z_i=h_i-h_{i+1}$ where
the height $h_i$ is a non-negative integer, with boundary condition
$h_{L+1}=0$. If the pile is in a stable state, a grain is dropped on
the site $i=1$. The update then proceeds for all sites in parallel. We
look for all sites which satisfy at least one of the two conditions
(i) it just toppled, (ii) its right-hand or left-hand neighbour toppled
\cite{ama_la_96}. If $i$ is such a site, it topples with probability 1
if $z_i>2$, with probability $\alpha\in[0,1]$ if $z_i=2$ and with
probability 0 if $z_i<2$. A toppling at the site $i$ means that $z_i$ is
decreased by 2 and $z_{i-1}$ and $z_{i+1}$ are increased by 1.

For $\alpha=0$ or $\alpha=1$ we recover the standard one-dimensional
BTW sandpile model  
with critical slope $z_c=1$ or $z_c=2$, respectively.
 In the intermediate region,
$0<\alpha<1$, self-organized criticality was found in numerical
simulations, with avalanche 
exponent $\tau=1.55\pm 0.02$ \cite{ama_la_97}. However, it is not
clear, what is the behaviour of the model for $\alpha$ close to either
1 or 0. It seems, that for a finite system the behaviour is
SOC (modified by finite size effects) only if $\alpha$ is not too
close to 1 or 0 \cite{be_be_ke_lo_mhi_99,markosova-unpublished}. The
behaviour of the 
system when the system size diverges and $\alpha$ stays close 0 or 1
was not clarified. We would like to study this question within the
approximation provided by a self-organized branching process. 

\subsection{Characteristic functions}
\label{sec:charfunction}

From the technical point of view we will use the method of
characteristic function (discrete Laplace transform), defined for a
function $f(s)$ on integer numbers $s$ as
$\hat{f}(\zeta)=\sum_{s=0}^\infty\,\zeta^s f(s)$. 

We will see that the
distribution of avalanches have generic form
\begin{equation}
P(s)\sim s^{-\tau}{\rm e}^{-s/s_0}
\label{eq:P-for-large-s}
\end{equation}
for large $s$. In the mean-field approximation or in the branching
process we have $\tau=3/2$, while in one-dimensional BTW sandpile the
exponent is $\tau=0$. The process is critical, if the cutoff avalanche
size $s_0$ diverges, $s_0\to\infty$.

In the language of characteristic functions the behaviour
(\ref{eq:P-for-large-s}) translates in the properties of the
singularity in $\hat{P}(\zeta)$.  
Generally we have $\hat{P}(\zeta)\sim
(\zeta-\zeta_0)^\eta + \;nonsingular\; part$. For the
one-dimensional BTW process we have $\eta=-1$, while true branching process has
$\eta=1/2$.
The cutoff is given by the distance of the singularity
from the point $\zeta=1$, namely $s_0\simeq 1/|\zeta_0-1|$. The
process is critical, if $\zeta_0=1$.

We will also see that the characteristic function for the branching
process is typically the solution of a quadratic equation. The
singular part of the characteristic function comes from the square
root of the discriminant $D(\zeta)$ of the equation,
i.e. $\hat{P}(\zeta)\sim 
\sqrt{D(\zeta)} + \;nonsingular\; part$. Therefore, $\eta=1/2$ and  
the cutoff is given by the solution of the equation
$D(\zeta_0)=0$. If $D(1)=0$, we have $s_0=\infty$ and the process is
critical.

\subsection{Branching process}
Let us first recall how the branching process is used to describe the
simplest case of the sandpile model, for  which in each toppling event
two grains are transferred to two randomly chosen nearest neighbours
(Manna model \cite{manna_91a}). There are $N_0$ sites in state $z=0$
and $N_1$ sites in state 
$z=1$. The branching process starts by
dropping a grain to a randomly chosen site. The probability of
becoming active (to topple) is $p=\frac{N_1}{ N_0+N_1}$. Two new
branches arise from an active site. Each of them is active with
probability $p$ and a tree is created iteratively. 
The branching process stops, when no active sites are
present at the end-points of the tree. The number of active sites, or
number of branchings, corresponds to the size of the avalanche. 
The probability distribution of avalanche sizes can be easily obtained
with the use of characteristic functions
\cite{za_la_sta_95,la_za_sta_96,ca_te_ste_96,ve_ma_ba_97,chu_ada_99}
and gives the mean-field value of the exponent $\tau=3/2$

Approximating the sand- or ricepile models by branching process is well
justified in high dimensions, where the activity returns to the same
point with very small probability. It seems, therefore, that the use
of branching processes in the opposite limit, in one dimension, lacks
sense, because the return of activity is very frequent.
However, we can use a very simple property  of the return of
activity to make the approximation sensible. Indeed, the most frequent
case when the activity returns to the 
same site is described by the following process.

If the site $i$ is active (it topples), a grain is transferred to site
$i+1$ which can become active. If that happens, another grain is
transferred back to site $i$ (and also to site $i+2$, but it is not
important now) and thus the site $i$ may become active again. This
observation 
leads to the modification of the branching process
suitable for the 
one-dimensional case. We should
take into account explicitly the return of the activity just in the
next step. We will do it by setting different branching probabilities
for site which was active just one step before (it is the site to the left)
and for the site which did not have to be (the site to the right).

Because the grains are added only on the site $i=1$, we have 
$z_i\ge 0\;\forall i$. The condition that the site topples with probability 1
if $z>2$ ensures that $z_i\le 2\;\forall i$. We denote $N_a$ number of
sites with $z=a$. So, picking randomly a site, we have probability 
$p_a=N_0/(N_0+N_1+N_2)$ of having $z=a$, where $a=0,1,2$.

Let us now describe the construction of the branching process 
corresponding to the
one-dimensional ricepile.  
There are three types of the points on the tree created by the
branching process, according to the value of $z\in\{0,1,2\}$. We
denote $q_a$ 
the probability that a point with $z=a$ do branch. The points with $z=0$
do not branch, i. e. $q_0=0$, while the points with $z=2$ always
branch, so $q_2=1$.  
The points with $z=1$ branch with probability
$\alpha$, i. e. $q_1=\alpha$. 
The approximation consists in
supposing that if a site did not topple in the previous step, 
it has probability $p_a$ of having $z=a$, while if the site did 
topple in the last step, the
probability of having $z=a$ is modified due to the previous toppling
to the value  
\begin{equation}
p^\prime_a=\frac{q_{a+1}\,p_{a+1}}{\sum_{b=0}^2q_{b+1}\,p_{b+1}}
\end{equation}
where we used $p_3=q_3=0$ for convenience.

\begin{figure}[ht]
\hspace*{10mm}
\includegraphics[scale=0.45]{\slaninafigdir/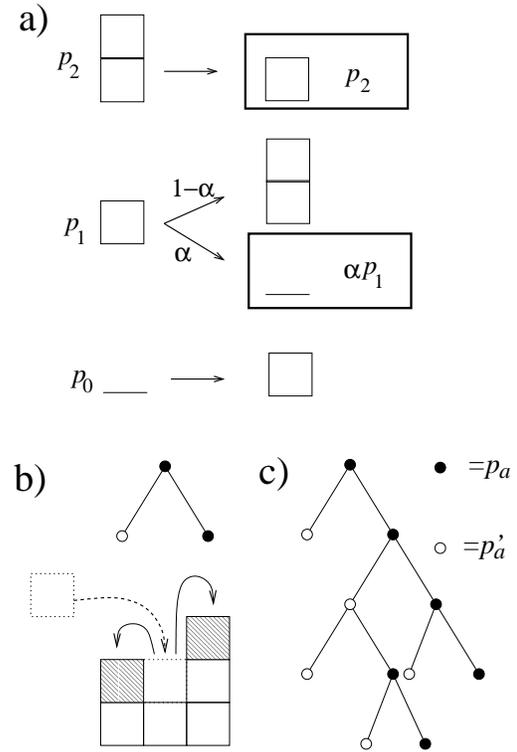}
\caption{Illustration of the branching process. 
In a) the processes following a grain drop are depicted.
Original configurations and their probabilities 
are in the left column, final ones in the
right column. The possible final configurations resulting from a
toppling are framed together with their non-normalized probabilities.
In b) the
correspondence is shown between one branching event and the toppling,
in which one new grain is added and two grains (shaded) are displaced
to the left and to the right from the toppling site. In c) a sample
realization of the tree is sketched. The full circles placed on the
right-hand branches correspond to
probabilities $p_a$, while empty circles on the left-hand branches
have modified probabilities $p'_a$.}  
\label{fig:branchingprocess}
\end{figure}

If a branching occurs at a site, two new branches (``left'' and
``right'') emanate from 
it. The probability that the 
right branch ends with a point with $z=a$ is $p_a$, while for the left
branch the probability is $p^\prime_a$.
This way the tree corresponding to the branching process is created.
The above described rules are illustrated in the
Fig. \ref{fig:branchingprocess}.

The root of the tree should be treated separately. The reason is that
in the ricepile model the avalanche starts by dropping
a grain always on the left edge of the pile, i. e. on the site
$i=1$. If it topples, it transfers a grain only to the 
right, while the grain going left falls off the system.
If we translated this feature to the description of our branching process,
the root 
would  consist either of a single non-branching point, or a point with
a single branch (the right one) emerging from it.  
However, we are interested in the regime of long trees, where the
different behaviour of the root from the rest of the tree is
irrelevant. So, we assume that in the branching process also the root
obeys the same rules as all other points. Thus, all points, including
the root, have either zero or two branches emanating from it.

The key quantity will be $P_n^a(s)$, the probability that a tree
consisting of $n$ 
levels starting with a point of type $a$ contains $s$ branchings.
The probability of having $s$ branchings (i. e. avalanche of size $s$)
is then $P_n(s)=\sum_a\; p_aP_n^a(s)$.
We can easily derive the recurrence relation for $P_n^a(s)$ which
becomes particularly simple if we use the characteristic function.
We obtain
\begin{equation}
\hat{P}_n^a(\zeta)=(1-q_a)+q_a \zeta \sum_{b,c=0}^2
p_b p^\prime_c
\hat{P}_{n-1}^b(\zeta)\hat{P}_{n-1}^c(\zeta)\;\; .
\label{eq:recurrence}
\end{equation}

A straightforward
calculation leads to the following 
equations for the characteristic
functions
\begin{equation}
\begin{split}
\hat{P}^0_n(\zeta)&=1\\
%
%
\hat{P}^1_n(\zeta)&=1-\alpha+\alpha\hat{P}^2_n(\zeta)
\label{eq:betweenp1andp2}
\end{split}
\end{equation}
and
\begin{equation}
P_n(s)=(\alpha p_1+p_2) P^2_n(s)\;\; {\rm for }\; s>0\;\; .
\end{equation}
Therefore the basic quantity of interest will be the characteristic
function $\hat{P}^2_n(\zeta)$. All properties of the branching process
can be computed from it. The set of equations (\ref{eq:recurrence})
thus represent a single recurrence equation for $\hat{P}^2_n(s)$,
which in the limit $n\to\infty$ 
 leads to quadratic equation
for the stationary distribution $\hat{P}^2(\zeta)=\lim_{n\to\infty}
\hat{P}^2_n(\zeta)$. We obtain explicitly 
\begin{equation}
\begin{split}
\frac{1}{\zeta}&\,\hat{P}^2(\zeta)=\label{eq:forP2}\\
& 
{\frac {\left (\alpha\,p_{{1}}+(1-\alpha)p_{{2}}\right )\left (
1-\alpha\,p_{{1}}-p_{{2}}\right )}{p_{{2}}+\alpha\,p_{{1}}}}
+
\\
&+
{\frac {\alpha\,p_{{2}}+2\,\alpha\,(1-\alpha)\,p_{{1}}p_{{2}}+(1-2\alpha)\,{p_{{2}}}^{2}
+{p_{{1}}}^{2}{\alpha}^{2}
}{p_{{2}}+\alpha\,p_{{1}}}}
\hat{P}^2(\zeta)\\
&+
\alpha\,p_{{2}}
(\hat{P}^2(\zeta))^2\quad .
\end{split}
\end{equation}

\subsection{Criticality}

 The discriminant $D(\zeta)$ of the equation (\ref{eq:forP2})
depends on the parameters $p_1$, $p_2$, and $\alpha$. The
branching process is critical if $D(1)=0$. This implies the following
relation  
\begin{equation}
-\alpha\,p_{{1}}-(1-\alpha)\,p_{{2}}+2\,\alpha\,p_{{1}}p_{{2}}+{p_{{1}}}^{2}{\alpha}^{2}+{p_{
{2}}}^{2}=0
\label{eq:condition}
\end{equation}
which determines a
surface in the parametric space. On this surface
the process is critical
and the distribution of avalanche sizes has a power-law tail with
exponent $\tau=3/2$.

However, the latter statement is not strictly true in the sense that if the
coefficient at the quadratic term in the equation (\ref{eq:forP2}) is
zero, the process is not a true branching process, because each parent
can have at most one offspring. It corresponds to a process with an
exponential distribution of avalanche sizes, which we will call, in
this work, a  
``one-dimensional BTW process''. The important feature which makes it
different from a generic branching process is that there are 
no true branching points. Indeed, there may be a non-zero probability that the
process stops at a given point, but there is zero probability to be
split into more than one branch. Therefore, the process does not
generate a tree-like structures, but linear chains of random length.
Both one-dimensional BTW and branching processes have the same general form
(\ref{eq:P-for-large-s})
of the distribution of avalanches for large $s$, but the
one-dimensional BTW process is characterised by the exponents
$\tau=0,\;\eta=-1$.
Therefore, together with checking the criticality condition
(\ref{eq:condition}) we must also look at the behaviour close to the
singularity.

We will prove in the section \ref{sec:fixedpoint} that in the
thermodynamic limit our 
ricepile model self-organizes so that the parameters stabilise at
values
\begin{equation}
\begin{split}
p_1=&\max(0,\frac{2\,\alpha-1}{\alpha})\\
p_2=&1-\alpha\;\;.
\end{split}
\label{eq:selforganizedp}
\end{equation}

If we insert these values into the criticality condition
(\ref{eq:condition}), we find that it is satisfied for any value of
$\alpha$, including the limit values of 0 and 1. At the same time we 
find that the  
singularity is always located at $\zeta_0=1$. (Indeed, as we discussed
in section \ref{sec:charfunction}, the criticality of the process is
equivalent to 
the condition $\zeta_0=1$.)
However, we find that
the type of the singularity corresponds to the exponents $\eta=1/2$,
$\tau=3/2$ (critical branching process) only for $\alpha$'s
within the open interval $(0,1)$, while at the points 0 and 1 the
model corresponds to one-dimensional BTW process. This can be easily
interpreted in the language of sand- and ricepiles. Indeed, for
$\alpha=0$ and $1$ the system recovers the behaviour of one-dimensional
BTW sandpile, which does not exhibit critical behaviour in the usual
sense. (In fact, the avalanche distribution {\it does} exhibit a
power-law distribution: all avalanche sizes have the same probability,
which corresponds to the power with exponent 0. But this is not the
situation we usually describe as critical behaviour.)

\subsection{Crossover behaviour}
\label{sec:crossover_behavior}

The question arises, how the behaviour with exponent $\tau=3/2$
inside the interval $[0,1]$ crosses over to the exponent $\tau=0$ 
at the edges. As the critical behaviour is related to the singularities
of the characteristic function, we will turn to the investigation of
the function $\hat{P}^2(\zeta)$ in more detail.

Indeed, we find that if we expand the solution of Eq. (\ref{eq:forP2})
for small values of the parameter $\rho$ defined as
\begin{equation}
\rho(\zeta)=
\frac{2\,\alpha\,(1-\alpha)}{\zeta^{-1}-1+2\,\alpha\,(1-\alpha)}
\label{eq:defrho}
\end{equation}
we can express the solution in terms of $\rho$ and expand in the
lowest order (for $\rho^2\ll 1$)  
\begin{equation}
\hat{P}^2(\zeta)=\frac{1}{\rho}-\sqrt{\frac{1}{\rho^2}-1}\simeq \frac{\rho(\zeta)}{2}\;\; .
\label{eq:P2approx}
\end{equation}
While, as noted earlier, the exact solution for $\hat{P}^2(\zeta)$
has always the singularity of the type $\eta=1/2$ for $\zeta\to\zeta_0=1$,
the approximate behaviour (\ref{eq:P2approx}) has a singularity with $\eta=-1$ 
located at the point $\zeta_0^\prime=(1-2\,\alpha\,(1-\alpha))^{-1}>1$.
When $\alpha$ goes to either 0 or 1, the value of $\zeta_0^\prime$
approaches 1. This suggests the following scenario. For large
avalanches, i. e. $1-\zeta\ll \zeta_0^\prime-1$ the singularity at
$\zeta_0=1$ is relevant and the avalanche size distribution has a
power-law tail with exponent $\tau=3/2$. 

For shorter avalanches,
i. e. $1-\zeta$ larger or comparable to $\zeta_0^\prime-1$ the
singularity at $\zeta_0^\prime$ becomes dominant.
Therefore, for short avalanches
we have one-dimensional BTW behaviour $P(s)\sim\exp(-s/s_0)$ with a
cutoff
\begin{equation}
s_0=|1-\zeta_0^\prime|^{-1}=
\frac{1-2\,\alpha\,(1-\alpha)}{2\,\alpha\,(1-\alpha)}\quad .
\label{eq:cutoff-full}
\end{equation}

The next step is investigation of the behaviour of $s_0$ when $\alpha$
approaches either 0 or 1. We find it by expanding the expression for
$\zeta_0^\prime$ as a function of $\alpha$ around the points 0 and 1,
respectively. To make the notation more compact, 
let us introduce the variable $\mu\in\{0,1\}$, which distinguishes the two
limit points $\alpha=0$ and 1.
We can see from (\ref{eq:cutoff-full}) that the cutoff diverges as 
\begin{equation}
s_{\rm 0}\simeq \frac{1}{2\,|\alpha-\mu|}
\label{eq:cutoff}
\end{equation}
for $\alpha\to\mu$.

On the
other hand, sufficiently close to the singularity at $\zeta\to\zeta_0=1$ the
exponent $\eta=1/2$ is relevant. The
question is, how close to $\zeta=1$ one behaviour crosses over to the
other. We have one-dimensional BTW behaviour for $\rho^2\ll 1$, while
critical branching process behaviour for $1-\rho^2\ll 1$. 
A typical crossover value $\zeta_{\rm cr}$ can be found by solving the
equation 
\begin{equation}
\rho(\zeta_{\rm cr})=\frac{1}{2}\quad .
\label{eq:forzetacr}
\end{equation}
The avalanche size distribution will exhibit the crossover
around $s_{\rm cr}=1/|1-\zeta_{\rm cr}|$. For $s\ll s_{\rm cr}$ the
one-dimensional BTW behaviour with exponential cutoff, diverging to
infinity for $\alpha=0$ and 1, will apply, while for $s\gg s_{\rm cr}$
the distribution will have power-law tail with usual mean-field
exponent $-3/2$, and therefore exhibits  self-organized criticality.

The point of the transition between SOC and one-dimensional BTW when
$\alpha$ approaches to 1 or 0 lies in the diverging crossover value for
the avalanche size. Similarly as in the case of $s_0$, by solving the
equation (\ref{eq:forzetacr}) with defintition (\ref{eq:defrho}) we find the
following limit behaviour 
\begin{equation}
s_{\rm cr}\simeq
\frac{1}{2\,|\alpha-\mu|}\simeq s_0 
\label{eq:crossover}
\end{equation}
for $\alpha\to\mu$.

We can see, comparing equations (\ref{eq:cutoff}) and
(\ref{eq:crossover}), that the cutoff for the one-dimensional BTW
behaviour is asymptotically equal to the crossover at which
the critical branching process behaviour sets on. This suggests the
scaling form
\begin{equation}
P(s)\simeq \frac{1}{s_0(\alpha)}\,F(\frac{s}{s_0(\alpha)})
\end{equation}
valid for $s\gg 1$ and $\alpha$ close to 0 and 1. 
The scaling function has the form
$F(x)\sim\,{\rm e}^{-x}$ for $x\ll 1$ and $F(x)\sim x^{-3/2}$
for $x\gg 1$. Indeed, we can find the Laplace transform of the scaling
function as
\begin{equation}
\int_0^\infty{\rm e}^{-x(y-1)}\,F(x)\,{\rm d}x =
y-\sqrt{y^2-1}\quad .
\end{equation}
From here we obtain immediately the expression for the scaling
function through the Bessel function of imaginary argument
\begin{equation}
F(x) = \frac{{\rm e}^{-x}}{x}\,I_1(x)\quad .
\end{equation}
The expected behaviour for $x\ll 1$ and $x\gg 1$ can be verified
directly by inspecting the asymptotic behaviour of the Bessel function.

\section{Self-organization}
\subsection{Self-organized branching process}

In the basic setup of our branching process, 
all three parameters $\alpha$, $p_1$, $p_2$ 
are freely chosen. However, in the ricepile model the only free parameter
is $\alpha$.  The number of sites with given
$z$ can change during an avalanche, so that also the probabilities $p_1$
and $p_2$ are modified. This defines a flow in the space
of parameters $p_1$, $p_2$. Our task now is to establish stable fixed
points of this dynamics and check whether they satisfy the condition
(\ref{eq:condition}). If that happens, we can conclude that the system
is self-organized critical.

There are four types of events, which can happen during an avalanche. 
Let us denote them $T2$, $T1$, $E1$, $E0$. 
In
the  
event $T2$, the point with $z=2$ receives a grain and topples. As a
result, the number of sites with $z=2$ is decreased by 1, $N_2\to
N_2-1$, and number of sites with $z=1$ is increased by 1, $N_1\to
N_1+1$. Similarly, in the event $T1$ point with $z=1$ topples, $N_1\to
N_1-1$ and $N_0\to N_0+1$, in event $E1$ site with $z=1$ receives a grain
but does not topple, $N_1\to N_1-1$ and $N_2\to N_2+1$, and finally in
event $E0$ site with $z=0$ receives a grain and does not topple,
$N_0\to N_0-1$ and $N_1\to N_1+1$.

Using the variables $y\in\{T,E\}$ and $a,b\in\{0,1,2\}$, let us denote
$s_{ayb,n}$ the number of events of the type $yb$ occurring at the
level $n$ within the branching process, on condition that the very
first site had $z=a$. There are
$s_{ayb}=\sum_{n=0}^\infty s_{ayb,n}$ such events in the entire
realisation of 
the branching process.
On average, there are $\langle s_{yb}\rangle =\sum_a\, p_a \langle
s_{ayb} \rangle$ events of the type $yb$.
The averages $\langle s_{yb}\rangle$ are of central importance for the
dynamics of the self-organization and can be easily obtained as follows.

For the characteristic
function of the probability
distribution%
of the number of events $s_{ayb,n}$ we obtain an equation analogical
to (\ref{eq:recurrence}). To study the self-organizat\-ion, we
will need only the average number of events, which is 
$\langle s_{ayb,n} \rangle$, calculated as derivative of the 
characteristic function.
Hence
\begin{equation}
\langle s_{ayb,n} \rangle
=q_a\sum_{c}(p_c+p_c^\prime)\langle s_{cyb,n-1} \rangle\quad .
\end{equation}
This is a set of three recurrence relations, which may be reduced
to one equation only, by considering the relations $\langle s_{0yb,n}
\rangle=0$ and $\langle s_{1yb,n} \rangle=\alpha\,\langle s_{2yb,n} \rangle$,
valid for $n>1$. If we take as the basic quantity the average $\langle
s_{2yb,n} \rangle$, we get a recurrence relation determining a
geometric sequence
\begin{equation}
\langle s_{2yb,n+1} \rangle = \kappa \langle s_{2yb,n} \rangle
\end{equation}
with quotient
\begin{equation}
\kappa =
\frac{\alpha\,p_2+(p_2 +\alpha\,p_1)^2}{p_2 +\alpha\,p_1}\quad .
\end{equation}
We recognise in the stationarity condition $\kappa=1$ the 
equation (\ref{eq:condition}), implying the criticality of the
branching process.

Summation of the infinite geometric series gives immediately
\begin{equation}
\langle s_{yb}\rangle
=\left(p_b+(p_b+p_b^\prime)\frac{\alpha\,p_1+p_2}{1-\kappa} 
\right)\langle s_{byb,1}\rangle
\label{eq:averages}
\end{equation}
where the initial conditions are given by $\langle
s_{bTb,1}\rangle=q_b$ and $\langle
s_{bEb,1}\rangle=1-q_b$.

The self-organization of the branching process is due to the changes
in the numbers $N_a$,  caused by the toppling (and non-toppling)
events. These numbers determine the probabilities $p_a$. Therefore,
for fixed $\alpha$  
the self-organized branching process (SOBP) ${\cal
S}(\alpha)$ 
consists of an (infinite) sequence of
branching processes
\begin{eqnarray}
&&{\cal S}(\alpha)=\\ 
&&=[{\cal B}(\alpha,p_1^{(0)},p_2^{(0)}), 
{\cal B}(\alpha,p_1^{(1)},p_2^{(1)}),
{\cal B}(\alpha,p_1^{(2)},p_2^{(2)}),
\ldots]\nonumber
\end{eqnarray}
where 
${\cal B}(\alpha,p_1,p_2)$ is the branching process determined by
fixed parameters $\alpha,p_1,p_2$, defined above. The branching
processes within the sequence differ only by the values of the
parameters $p_1$, and $p_2$.
Let us consider the $t$-th branching process in the sequence. When
realised, it changes the original values of the numbers $N_a$, or,
equivalently, the values of the parameters $p_a$. The average 
change is uniquely determined by the average number of events 
$\langle s_{yb}\rangle$ .
So, the 
SOBP is entirely determined by the transition relations
connecting the values of the parameters in the $t$-th and $t+1$-st
step
\begin{equation}
p_i^{(t+1)}-p_i^{(t)}=T_i(p_1^{(t)},p_2^{(t)})
\label{eq:evolution}
\end{equation}
for $i\in\{1,2\}$.
We find explicitly
\begin{equation}
\begin{split}
&T_1(p_1,p_2)=\\
&\frac{\alpha p_1+(1-\alpha) p_2 -\alpha (2 -\alpha) p_1^2 -p_2^2 -2(1- \alpha) p_1 p_2
}{\alpha p_1+(1-\alpha) p_2 +2 \alpha p_1 p_2
+p_2^2+\alpha^2 p_1^2}\\    
&T_2(p_1,p_2)=\\
&\qquad\frac{ \alpha(1-\alpha) p_1^2+(1-2 \alpha)p_1p_2 
}{\alpha p_1+(1-\alpha) p_2 +2 \alpha p_1 p_2  +p_2^2+\alpha^2 p_1^2}\quad .
\end{split}
\label{eq:evolutionofp}
\end{equation}

\subsection{Fixed point}
\label{sec:fixedpoint}

The fixed point of the self-organization dynamics 
can be found immediately by equating the right-hand sides of equations
(\ref{eq:evolutionofp}) to zero. 
Direct solution of the two coupled equations gives three fixed points
\begin{eqnarray}
p_1=&0,\;&p_2=0\label{eq:fixed1}\\
p_1=&0,\;&p_2=1-\alpha\label{eq:fixed2}\\
p_1=&\frac{2\alpha-1}{\alpha},\;&p_2=1-\alpha\quad .
\label{eq:fixed3}
\end{eqnarray}
The correct solution is determined by stability considerations. The
relations (\ref{eq:evolutionofp}) are linearised around the fixed
points and the eigenvalues of the resulting matrices of rank 2 are
found. The result is that the fixed point (\ref{eq:fixed1}) is always
unstable, while (\ref{eq:fixed2}) is stable for $\alpha\in [0,1/2)$
and (\ref{eq:fixed3}) is stable for $\alpha\in (1/2,1]$. For $\alpha =
1/2$ the fixed points (\ref{eq:fixed2}) and (\ref{eq:fixed3}) coincide
and both of them are marginally stable (i. e. the eigenvalues have
zero real part).
 
Therefore, we find that the fixed point corresponds to the
values of the probabilities
\begin{equation}
\begin{split}
p_1=&\max\,(0,\frac{2\,\alpha-1}{\alpha})\\
p_2=&1-\alpha
\end{split}
\end{equation}
which proves the already announced result of
Eq.(\ref{eq:selforganizedp}).

\subsection{Finite-size effects}

In the numerical simulations of the ricepile model \cite{be_be_mhi_zha_99,be_be_ke_lo_mhi_99,markosova_00,markosova-unpublished}
attention is paid to the fact that the critical behaviour is observed
only for large enough systems with $\alpha$ not too close to neither 0
nor 1. We have already shown how the crossover length blows up when
$\alpha$ approaches the edge values 0 or 1. It is obvious then, that
for small systems the crossover value of the avalanche size 
may not be accessible and the critical regime in the tail of the
distribution is not observed at all.
In this subsection we will investigate the consequences of finite length
of the branching process. There are two phenomena where the finite
size enters the problem. First, if the maximum number of generations in the
branching process is $L$, instead of infinity, the distribution of the
avalanche sizes will not extend to infinity either, but will be
bounded by $s<s_{\rm max}=2^L-1$. Moreover, if we take for example
$p_1=1$, $p_2=0$, $\alpha=1$, then all avalanches will have size $L$,
therefore a peak at $s=L$ will occur, 
$P(s)=\delta(s-L)$. If we move slightly from this position
by increasing $p_2$ and decreasing $p_1$ and $\alpha$, a
structure of multiple peaks located at $s=L,2L-1,3L-3,...$ will
appear. This makes the analysis very complicated.

Second consequence is the shift in the self-organized value of the
parameters $p_1$ and $p_2$, which for finite $L$ will deviate from the
critical values. Therefore, the avalanche-size distribution will
develop an exponential cutoff in the form
$P(s)\propto s^{-3/2}\exp(-s/s_1)$.

As the first problem brings new particular difficulties, we will
concentrate only on the second one. This makes the analysis less
consistent, but feasible. Thus, we should stress that in the following
we will suppose that the branching process in question has unbounded
length, but the self-organization is made in such a way, that only the
first $L$ generations of the branching process are taken into account.

Instead of working with finite-$L$ version of the equations
(\ref{eq:evolution}) and 
(\ref{eq:evolutionofp}), describing the approach to the fixed point,
we can use the set of equations
\begin{equation}
\begin{split}
\langle
s_{E1}\rangle\,=&\,\langle s_{T2}\rangle\\
\langle
s_{E0}\rangle\,=&\,\langle s_{T1}\rangle
\end{split}
\label{eq:equilibrium}
\end{equation}
which determine the position of the fixed point. The only information
lost in Eq. (\ref{eq:equilibrium}) is the stability of the fixed
points. However, we suppose the stability will not be
affected by the finite-size effects. Therefore, we will rely on the
stability analysis performed for $L=\infty$ also in the case of finite
$L$ and calculate the finite-size corrections starting with  
Eq. (\ref{eq:equilibrium}).

The point is that the equations (\ref{eq:equilibrium}) should hold
also for finite $L$. In fact, the expression (\ref{eq:averages}) for
the averages $\langle s_{yb}\rangle$  
assume the same form, only
the factor $(\kappa - 1)$ arising in the $L=\infty$ version should be
replaced by the factor $K=(\kappa - 1)/(\kappa^{L-1}-1)$. Assuming $K$
small for large $L$, we can find $p_1$ and $p_2$ in lowest order in
$K$. Then, we return to the definition of $K$ and find that $K\propto
L^{-1}$, confirming that our approach is consistent.

Hence,
for finite $L$ we find, by solving the
equations (\ref{eq:equilibrium}) to lowest order in $1/L$,
for $\alpha\in(0,1/2)$
\begin{equation}
\begin{split}
p_1 &=
-
\frac{1-\alpha}{(2\alpha-1)^2}\;\frac{\ln(1-\gamma)}{L} 
+ O(\frac{1}{L^2})\\
p_2 &= 1-\alpha -
\frac{1-\alpha}{2\alpha-1}\;\frac{\ln(1-\gamma)}{L} 
+ O(\frac{1}{L^2})
\end{split}
\end{equation}
and for $\alpha\in(1/2,1)$
\begin{equation}
\begin{split}
p_1 &=  \frac{2\alpha-1}{\alpha}+ 
\frac{5\alpha^2-5\alpha+1}{(2\alpha-1)^2\,\alpha}\;\frac{\ln(1-\gamma)}{L} 
+ O(\frac{1}{L^2})\\
p_2 &= 1-\alpha 
+
\frac{1-\alpha}{2\alpha-1}\;\frac{\ln(1-\gamma)}{L} 
+ O(\frac{1}{L^2})
\end{split}
\end{equation}
where we denoted
\begin{equation}
\begin{split}
\gamma&=
\frac{1}{2}\frac{1-2\alpha}{1-\alpha}\quad{\rm for}\quad
\alpha\in(0,1/2)\\ 
\gamma&=\frac{1}{2}\frac{2\alpha-1}{\alpha}\quad{\rm for}\quad
\alpha\in(1/2,1)
\quad. 
\end{split}
\end{equation}
The above formulae confirm that the explicit limit $L\to\infty$ gives
the same result as obtained previously when working directly with
$L=\infty$. 

Using these results we can find the position of the square-root
singularity in the characteristic function for the avalanche size
distribution, solving the equation\linebreak $D(\zeta_0)=0$. The distance from
1 then determines the exponential cutoff of the distribution.  We find
\begin{equation}
1/s_1 = |\zeta_0-1| = \frac{\sigma(\alpha)}{L^2}
+O(\frac{1}{L^3})
\end{equation}
where
\begin{equation}
\sigma(\alpha)=\frac{\ln^2(1-\gamma)}{
4\,\alpha\,(1-\alpha)}\quad{\rm for}\quad
\alpha\in(0,1)
\end{equation}
and asymptotically for $L\to\infty$ and $\alpha$ fixed 
the avalanche distribution
becomes the function of $s\,L^{-2}$ only, 
\begin{equation}
P(s;\alpha,L)\propto L^{-3}\,G(sL^{-2}\,\sigma(\alpha))
\end{equation}
and he scaling function has the form
\begin{equation}
G(x)=x^{-3/2}{\rm e}^{-x}\quad .
\end{equation}
This scaling holds well for all $\alpha$ with exception of the point
$\alpha=1/2$, where we have $\gamma=0$ and hence $\sigma(\alpha) = 0$. 
Then, the next order in $1/L$ takes over and the scaling changes.

Let us use again the variable $\mu\in\{0,1\}$, which distinguishes the two
limit points $\alpha=0$ and 1.
The factor $\sigma(\alpha)$
diverges as $\sigma(\alpha)\simeq \sigma_0\,|\alpha-\mu|^{-1}$ for $\alpha\to
\mu$, where $\sigma_0=(\ln 2)^2/4$.
Therefore, we can write the following scaling form for the avalanche
size distribution 
\begin{equation}
P(s;\alpha,L)\propto L^{-3}|\alpha-\mu|^{-\frac{3}{2}}\,G(s\,L^{-2}|\alpha-\mu|^{-1}\,\sigma_0)
\label{eq:scaling}
\end{equation}
for
$\alpha\to \mu$.

We can see that the power-law distribution holds only for avalanches
shorter than $L^2\,|\alpha-\mu|$. In other words, if the parameter
$\alpha$ is close to the end-points of the interval $[0,1]$,
we need to have systems of the size $L\gg 1/\sqrt{|\alpha-\mu|}$ for being
able to observe any sign of self-organized criticality.

In the above calculations we tacitly assumed that we beyond  the regime
we called ``one-dimensional BTW'' in the section
\ref{sec:crossover_behavior}. This means $s\gg s_{\rm cr}$. In fact,
we can always reach this regime by choosing $L$ large
enough. Therefore the presence of the one-dimensional regime does not
influence the scaling behaviour for large $L$. More precisely, we
should have $ L^2\,|\alpha-\mu| \gg s_{\rm cr}$. But because $s_{\rm
cr}$ itself diverge for $\alpha\to \mu$ as $|\alpha-\mu|^{-1}$, we obtain
a stronger condition for the scaling (\ref{eq:scaling}) to be valid, namely
\begin{equation}
L\gg |\alpha-\mu|^{-1}
\end{equation}
if $\alpha\to \mu$.

\section{Conclusions}

We investigated analytically the self-organized critical ricepile
model.
We defined
a self-org\-aniz\-ed
branching process, suitable for one-dimensional pr\-oblems.
 The model is characterised by
the parameter $\alpha\in [0,1]$, 
the probability of toppling at a sub-threshold
site. For both limit values $\alpha=0$ and $\alpha=1$ the model is
equivalent to the one-dimensional BTW model with trivial (uniform)
distribution of avalanches.

We found that in the thermodynamic limit the system is self-organized 
critical for all values of $\alpha$ within the open interval
$(0,1)$, with power-law tail in the distribution of avalanche sizes
with mean-field 
value of the exponent, $\tau=\frac{2}{3}$. However, the power-law
behaviour holds only for avalanches longer than a certain crossover value
of the avalanche size. The crossover diverges when $\alpha$ approaches to
either of the limit points of the interval $[0,1]$. We found the
scaling as well as the exact form of the scaling function for
avalanche distribution close to these limit points.
This describes how the one-dimensional BTW behaviour develops when
approaching the limit points.

The finite-size effects play important role in determining
whether the model is self-organized critical or not. In our model the
SOC behaviour starts to occur at the larger sizes the closer we are to
the limit points $\alpha=0$ or $1$. We found  
the form of the finite size scaling in our self-organized branching
process and determined the necessary condition for the 
the power-law regime in the avalanche distribution to be observable, when we
approach to the  limit points.

\begin{acknowledgement}
I wish to thank M\'aria Marko\v{s}ov\'a for numerous useful
discussions which motivated me to this work. I am indebted to  Petr
Chvosta for clarifying comments 
regarding stochastic processes. 
This work was partially supported by the Grant
Agency of the Czech Republic, project No. 202/00/1187.%
\end{acknowledgement}
{}
\end{document}